\UseRawInputEncoding
\documentclass[a4paper,11pt]{article}
\pdfoutput=1 % if your are submitting a pdflatex (i.e. if you have
\usepackage{jcappub} % for details on the use of the package, please
\usepackage[T1]{fontenc} % if needed
\title{Nonminimal Scalar-curvature Coupling Noncanonical Warm Inflation}

\author[a,*]{Zhi-Peng Peng,\note[*]{Corresponding author.}}
\author[a]{Shi-Chao Zhang,}
\author[a]{Xiao-Pei Xu,}
\author[a]{Ruan-Jing Zhang,}
\author[b,*]{Xiao-Min Zhang,}
\author[c,*]{Xi-Bin Li}

\affiliation[a]{School of Physics and Advanced Energy, Henan University of Technology, No.100, Lianhua Street, Zhengzhou, China}
\affiliation[b]{School of Science, Qingdao University of Technology, No.777, Jialingjiang East Road, Qingdao, China}
\affiliation[c]{College of Physics and Electronic Information, Inner Mongolia Normal University, No.81 Zhaowuda Road, Hohhot, China}

\emailAdd{zhipeng@haut.edu.cn}
\emailAdd{2655801626@qq.com}
\emailAdd{xuxiaopei@haut.edu.cn}
\emailAdd{zhangruanjing@126.com}
\emailAdd{zhangxm@mail.bnu.edu.cn}
\emailAdd{lxbimnu@imnu.edu.cn}

\abstract{We investigate a noncanonical warm inflationary model with a nonminimal coupling between the inflaton field and scalar curvature. Within the Jordan frame, we derive the modified background equations and establish the slow-roll conditions through a linear stability analysis. We find that the nonminimal coupling enhances the effective damping and modifies the potential gradient, facilitating slow-roll evolution even for relatively steep potentials. Using the stochastic approach, we derive the Langevin equation for inflaton fluctuations and obtain the primordial scalar and tensor power spectra. Our results show that the scalar spectrum is nearly scale-invariant, while the combined effects of nonminimal coupling, noncanonical dynamics, and thermal dissipation can further suppress the tensor-to-scalar ratio. Moreover, the model can sustain warm inflation even in the weak-dissipation regime and accommodate sub-Planckian field excursions in suitable parameter regions. These findings highlight the potential of this framework as a viable extension of conventional warm inflation.}

\begin{document}
\maketitle
%\flushbottom

\section{Introduction}
\label{sec:intro}

Inflation, a period of quasi-exponential accelerated expansion in the early Universe, provides an effective framework for addressing the horizon, flatness, and monopole problems of the standard hot Big Bang cosmology \cite{Guth1981,Linde1982,AlbrechtSteinhardt1982}. It also offers a mechanism for generating the primordial perturbations that seed the formation of large-scale structure and produce the temperature anisotropies observed in the cosmic microwave background (CMB) \cite{Planck2018,Planck:2019kim}. In the standard cold inflation picture, the inflaton evolves in an approximately vacuum-dominated background, and its quantum fluctuations are stretched beyond the Hubble radius. A subsequent reheating stage is then required to transfer the inflaton energy into radiation. Warm inflation provides an alternative picture in which interactions between the inflaton and other fields continuously produce radiation during inflation \cite{Berera1995,BereraFang1995}. The radiation energy density can remain subdominant while the Universe is inflating, allowing accelerated expansion and a thermal bath to coexist. In this scenario, the inflaton equation of motion contains a dissipative term \(\Gamma\dot{\phi}\), where \(\Gamma\) is the dissipation coefficient, and the radiation bath can contribute thermal fluctuations to the primordial density perturbations \cite{Berera1995,Berera2000,Hall:2003zp}. Depending on the temperature and dissipation regime, these thermal fluctuations can significantly affect the scalar power spectrum. The additional damping and radiation production also modify the background evolution and the conditions required for slow roll \cite{Moss:2008yb,delCampo:2010by}. Warm inflation can therefore connect inflationary dynamics more directly to particle interactions and, in suitable models, provide a smooth transition toward the radiation-dominated era without a separate, abrupt reheating stage. Many problems suffered in standard inflation such as $\eta$ problem \cite{Dine:1995uk,Kolda:1998kc} and the overlarge amplitude of the inflaton \cite{Berera:2006xq,Berera:2008ar} can be cured in warm  inflation.

Inflation is commonly described by a canonical scalar field with a Lagrangian density \(\mathcal{L}=X-V(\phi)\), where \(X\equiv-\frac{1}{2}g^{\mu\nu}\partial_{\mu}\phi\,\partial_{\nu}\phi\) for the \((-+++)\) metric convention. Beyond this canonical description, a broad class of models replaces the standard kinetic term with a nonlinear function \(\mathcal{L}(X,\phi)\). Early studies of \(k\)-inflation showed that nonlinear kinetic terms can drive accelerated expansion, while subsequent analyses of cosmological perturbations established that the noncanonical kinetic structure can modify the sound speed and the primordial perturbation spectra \cite{Armendariz-Picon:1999hyi,Garriga:1999vw}. In warm inflation, the combination of noncanonical kinetics and thermal dissipation therefore provides a natural extension of canonical warm inflation. Warm DBI inflation was studied as a concrete example in which the sound speed and dissipation strength jointly affect the scalar spectrum and non-Gaussianity \cite{Cai:2010wt}. Our previous work subsequently formulated a general noncanonical warm inflation framework, deriving its background equations, slow-roll conditions, linear stability criteria, and an approximate scalar power spectrum \cite{ZhangZhu2014,Li:2018riw}. And other developments include warm tachyon inflation \cite{Zhang:2013waa} and warm \(k\)-inflation \cite{Peng:2016yvb,Peng:2018bir,Shen:2023ssx}. These studies show that noncanonical warm inflation can produce dynamics and perturbation spectra that differ from those of canonical warm inflation; in particular, the kinetic structure and thermal effects can either reinforce or compete with one another. The quantitative consequences, including any changes to the inflationary energy scale or tensor-to-scalar ratio, are model-dependent and must be determined for each choice of \(\mathcal{L}(X,\phi)\), dissipation coefficient, and background solution.

Most inflationary models have been constructed within the framework of general relativity (GR). Extending inflation to the framework of nonminimal coupling theory broadens its scope. Nonminimal couplings between the inflaton and gravity can be broadly classified into two types. In nonminimal scalar--curvature coupling (NMSC) models, the inflaton couples directly to the Ricci scalar through a term of the form \(f(\phi)R\) \cite{spokoiny1984inflation,FutamaseMaeda1989,Kaiser:1994vs,BezrukovShaposhnikov2008}. In nonminimal derivative coupling (NMDC) models, derivative terms of the inflaton are instead coupled to curvature tensors, commonly through a coupling to the Einstein tensor \cite{Germani:2010gm,Dalianis:2019vit,Karydas:2021wmx}.  Early studies investigated how nonminimal derivative coupling modifies the dissipative dynamics and cosmological perturbations of warm inflation, including its effects on the slow-roll conditions, scalar power spectrum, and non-Gaussianity \cite{NozariShoukraniRashidi2014,SadjadiGoodarzi2015}. More recently, noncanonical warm inflation has been formulated in the NMDC framework, in which the inflaton kinetic term couples to the Einstein tensor. The combined gravitational and thermal damping in this scenario leads to modified slow-roll and stability conditions, and its effects on the primordial power spectrum and related observables have been analyzed \cite{ZhangEtAl2025NMDC}. Warm inflation with a direct scalar--curvature coupling \(f(\phi)R\) has also been studied in some specific canonical models. Examples include nonminimally coupled warm inflation with a quantum-corrected self-interacting potential and nonminimally coupled warm Higgs inflation in both metric and Palatini formulations \cite{SamartEtAl2022,EadkhongEtAl2023}. These studies have examined the background dynamics and observational predictions of scalar--curvature coupled warm inflation, although the analyses focus on particular inflaton sectors and dissipative regimes.

Motivated by these developments, we will conduct a systematic analysis of noncanonical warm inflationary models incorporating a scalar--curvature coupling \(f(\phi)R\) in this paper. This framework makes it possible to study systematically how these ingredients jointly affect the slow-roll evolution, linear stability, radiation production, and primordial perturbation spectra. We firstly derive the background equations and their slow-roll approximation, and then examine the conditions under which the inflationary solution is a stable attractor. We then study the scalar and tensor perturbations and their implications for the primordial power spectra and inflationary observables. In particular, the analysis tracks how the noncanonical kinetic structure, thermal dissipation, and the field dependence of the effective gravitational coupling jointly affect the background evolution and perturbation amplitudes. This framework provides a basis for comparing the model with canonical and noncanonical warm inflation in general relativity.

The paper is organized as follows. Section~\ref{sec:model} presents the model and derives its background equations. Section~\ref{sec:slowroll} discusses the slow-roll regime and the linear stability conditions. Section~\ref{sec:perturbations} studies cosmological perturbations and the resulting observables. Section~\ref{sec:conclusion} summarizes our results.

\section{Slow-roll regime of scalar-curvature coupling noncanonical warm inflation}
\label{sec:model}

Following the framework of Spokoiny \cite{spokoiny1984inflation}, we consider inflation in a broken-symmetric theory of gravity, in which the effective gravitational coupling is generated dynamically through the nonminimal interaction between a scalar field and the Ricci curvature. Considering all components of the universe, the total action in the NMSC noncannonical warm inflationary scenario under the Jordan frame is expressed as follows:
\begin{equation}
  \label{1-action} S = \int d^4 x \sqrt{- g} \left[ \frac{1}{2} M_p^2 f (\phi)
  R + \mathcal{L} (X, \phi) + \mathcal{L}_{\text{rad}} +
  \mathcal{L}_{\text{int}} \right],
\end{equation}
where $g$ is the determinant of the metric tensor $g_{\mu\nu}$, $M_p^2 \equiv (8 \pi G)^{- 1}$ is the reduced squared Planck mass with $G$ being the gravitational constant, the functional $f(\phi)$ is a generic function of the scalar field $\phi$ coupled to the scalar curvature $R$, and the standard kinetic term $X \equiv - \frac{1}{2} g^{\mu \nu} \partial_{\mu} \phi \partial_{\nu} \phi$. We focus on the positive-coupling branch, the coupling parameter \(\xi>0\), for which the nonminimal coupling function \(f(\phi)=1+\xi\phi^2/M_{p}^2\) remains larger than unity during inflation \cite{BezrukovShaposhnikov2008}. Also, the Lagrangian density for the noncanical inflaton field is $\mathcal{L}_{\text{noncan}} = \mathcal{L} (X,\phi)$, a general function of the field and the kinetic term $X$, and for simplicity, $\mathcal{L}$ is used to denote $\mathcal{L} (X, \phi)$ hereinafter. $\mathcal{L}_{\text{rad}}$ describes the radiation fields, while $\mathcal{L}_{\text{int}}$ accounts for the interactions between the inflaton and all other subdominant components.

A physically viable noncanonical Lagrangian density is required to satisfy \(\mathcal{L}_{X}\geq 0\) and \(\mathcal{L}_{XX}\geq 0\), where the subscript \(X\) denotes differentiation with respect to \(X\). By contrast, the subscripts in \(\mathcal{L}_{g}\), \(\mathcal{L}_{R}\), and \(\mathcal{L}_{\text{int}}\) are merely labels. These conditions ensure consistency with the null energy condition and support the physical propagation of perturbations \cite{Bean:2008ga,Franche:2009gk}. Together with the field-normalization condition, they yield \(\mathcal{L}_{X}\geq 1\). Varying the action in Eq.~(\ref{1-action}) with respect to the scalar field then gives the inflaton equation of motion:
\begin{equation}
  \frac{\partial [\mathcal{L} (X, \phi) + \frac{1}{2} M_p^2 f (\phi) R +
  \mathcal{L}_{\text{int}}]}{\partial \phi} - \left( \frac{1}{\sqrt{- g}}
  \right) \partial_{\mu} \left[ \sqrt{- g} \frac{\partial \mathcal{L} (X,
  \phi)}{\partial (\partial_{\mu} \phi)} \right] = 0.
\end{equation}
In the spatially flat Friedmann-Lemaitre-Robertson-Walker(FLRW) universe with scale factor $a(t)$ where $t$ is the cosmic time, $R = 6 \left( \frac{\ddot{a}}{a} + H^2 \right) = 6 (2 H^2 + \dot{H})$ where the Hubble parameter $H=\frac{\dot{a}}{a}$ and the inflaton mean field is assumed to be homogeneous, i.e. $\phi = \phi (t)$. Hence, the motion equation for the inflaton field reduces to
\begin{eqnarray}
  \label{EOM} \left[ \frac{\partial \mathcal{L} (X, \phi)}{\partial X} + 2 X\frac{\partial^2 \mathcal{ L} (X, \phi)}{\partial X^2} \right] \ddot{\phi} & + &\left( 3 H \frac{\partial \mathcal{L} (X, \phi)}{\partial X} +
  \frac{\partial^2 \mathcal{L} (X, \phi)}{\partial X \partial \phi} \dot{\phi}\right) \dot{\phi}  \nonumber\\
  &-& 3 M_p^2 f_{\phi} (2 H^2 + \dot{H})  -\frac{\partial [\mathcal{L} (X, \phi) + \mathcal{L}_{\text{int}}]}{\partial \phi} = 0,
\end{eqnarray}
where $X = \frac{1}{2} \dot{\phi}^2$ in the FLRW universe.

Assuming that the Lagrangian density separates into kinetic and potential contributions, $\mathcal{L}(X,\phi)=K(X)-V(\phi)$, we take the noncanonical kinetic function \(K(X)\) to depend weakly, or not at all, on \(\phi\) \cite{Franche:2009gk}. In the case where \(K\) depends only on \(X\), it follows that \(\mathcal{L}_{X\phi}=0\) and \(K_X=\mathcal{L}_X\). The term \(\Gamma\dot{\phi}\) represents the dissipation of the inflaton into other fields and acts as a thermal damping term \cite{Berera1995,BereraFang1995}. The terms in \(\partial\mathcal{L}(X,\phi)/\partial\phi\) and \(\partial\mathcal{L}_{\mathrm{int}}/\partial\phi\) in Eq.~(\ref{EOM}) that are independent of \(\dot{\phi}\) can be combined into an effective potential \(V_{\mathrm{eff}}\). This effective potential incorporates thermal corrections and depends on both the inflaton field and the temperature. The temperature is that of the radiation bath, which remains nonzero because of the inflaton's dissipative energy transfer, provided that the temperature dependence of the dissipation coefficient satisfies the condition derived in Eq.~(\ref{c}) of Section 3. With these assumptions, the equation of motion can be rewritten as follows:
\begin{equation}
   \label{Inflaton} \mathcal{L}_X c_s^{- 2} \ddot{\phi} + (3 H \mathcal{L}_X + \Gamma) \dot{\phi}- 3 M_p^2 f_{\phi} (2 H^2 + \dot{H}) + V_{\text{eff}, \phi}(\phi,T) = 0,
\end{equation}
where the sound speed $c_s$ is an important parameter for the noncanonical field, which characterizes the propagation speed of scalar perturbations: $c_{s}^{2}=\frac{p_{X}(\phi,X)}{\rho_{X}(\phi,X)}=\frac{\mathcal{L}_{X}}{\mathcal{L}_{X}+2X\mathcal{L}_{XX}}$ where the subscript $X$ denotes a derivative. For simplicity, the thermal effective potential $V_{\text{eff}}(\phi,T)$ is denoted as $V(\phi,T)$ hereinafter, and the subscript $\phi$ denotes a derivative.

By varying the action (\ref{1-action}) with respect to the metric, the field equations for the metric are obtained as follows
\begin{equation}
  \label{GR} f (\phi) G_{\mu \nu} = \frac{1}{M_p^2} T_{\mu \nu},
\end{equation}
where the total energy--momentum tensor $T_{\mu \nu}$ containing the inflaton and radiation field contributions is
\begin{equation}
  T_{\mu \nu} = \nabla_{\mu} \phi \nabla_{\nu} \phi \mathcal{L}_X + g_{\mu
  \nu} \mathcal{} \mathcal{L} (X, \phi) - M_p^2 (g_{\mu \nu} \Box f - \nabla_{\mu}
  \nabla_{\nu} f)+ T s u_{\mu} u_{\nu},
\end{equation}
where the d'Alembert operator is expressed as $\Box = \frac{1}{\sqrt{- g}} \partial_{\nu} \left( \sqrt{- g} g^{\mu \nu} \partial_{\mu} \right)$, $T$ is the universal temperature, $s$ is the entropy density and $u_{\mu}=(-1,0,0,0)$ is the radiation fluid 4-velocity vector. The energy-momentum tensor has  the form $T^{\mu}_{\nu}=\text{diag} (-\rho,p,p,p)$ with
\begin{equation}
  \rho = 2 X \mathcal{L}_X - K (X) + V (\phi, T) - 3 M_p^2 H f_{\phi} \dot{\phi} + T s,
\end{equation}
\begin{equation}
  p = K (X) - V(\phi, T)+ M_p^2 [2 H f_{\phi} \dot{\phi} + (f_{\phi} \ddot{\phi} + f_{\phi \phi} \dot{\phi}^2)],
\end{equation}
where $\rho$ and $p$ are the total energy density and pressure of the system, respectively. Here, $\rho$ and $p$ have an explicit dependence on the Hubble rate, therefore, the gravitational field equations are given by
\begin{equation}
 \label{FRW1} 3 f H^2 = \frac{1}{M_p^2} \rho = \frac{1}{M_p^2} [2 X \mathcal{L}_X - \mathcal{L} - 3 M_p^2 H f_{\phi} \dot{\phi} + T s],
\end{equation}
and
\begin{equation}
\label{FRW2}  2 f \dot{H} = - \frac{1}{M_p^2} (\rho + p) = - \frac{1}{M_p^2} [2 X \mathcal{L}_X + M_p^2 (f_{\phi} \ddot{\phi} + f_{\phi \phi} \dot{\phi}^2 - H f_{\phi} \dot{\phi}) + T s],
\end{equation}
Now, considering the thermodynamic relation $U= F+ T s$ where $U$ is the total energy and $F$ is total free energy, the free energy density will be
\begin{equation}
  f_{\text{free}} = 2 X \mathcal{L}_X - K (X) + V (\phi, T) - 3 M_p^2 H f_{\phi} \dot{\phi}.
\end{equation}
Indeed, through the definition of entropy in thermodynamics, $s(\phi,T)$ has following relation with potential
\begin{equation}
  \label{thermal} s = - \frac{\partial f_{\text{free}}}{\partial T} = -V_{T}(\phi,T).
\end{equation}

Let us rewrite the field equations (\ref{GR}) as
\begin{equation}
  G_{\mu \nu} = \frac{1}{M_p^2} \tilde{T}_{\mu \nu},
\end{equation}
where $\tilde{T}_{\mu \nu} \equiv T_{\mu \nu} / f$ and the conservation law of the energy--momentum tensor $\nabla^{\mu} \tilde{T}_{\mu \nu} = 0$ is valid as a direct consequence of the Bianchi identities $\nabla^{\mu} G_{\mu \nu} = 0$. So the total energy conservation equation is
\begin{equation}
  \label{Energy} \dot{\rho} + 3 H (\rho + p) = \frac{\dot{f}}{f} \rho,
\end{equation}
where the term $\frac{\dot f}{f}\rho$ acts as an effective source term induced by the time dependence of the nonminimal coupling function \(f(\phi)\). Physically, this term represents an effective exchange of energy between the scalar-field sector and the gravitational sector caused by the evolving nonminimal coupling. Therefore, it should not be interpreted as a violation of total energy-momentum conservation, since the rescaled energy-momentum tensor \(\widetilde T_{\mu\nu}=T_{\mu\nu}/f\) remains covariantly conserved.

Combining Eqs.(\ref{Inflaton}) and (\ref{Energy}), we can get the entropy production equation:
\begin{equation}
  \label{entropy} T \dot{s} + 3 H T s = \Gamma \dot{\phi}^2.
\end{equation}
When thermal correction to field masses is negligible (the conditions for their being negligible during the slow-roll epoch will be given in the next section), the effective potential can be written as $V(\phi, T) \simeq V(\phi) +V(T)$ and therefore $V(T)$ contributes to the energy density of the radiation field resulting in $\rho_r = 3 T s/4 $ and Eq.(\ref{entropy}) is equivalent to the radiation energy density producing equation:
\begin{equation}
  \dot{\rho}_r + 4 H  {\rho}_r = \Gamma \dot{\phi}^2.
\end{equation}

Substituting Eqs.(\ref{FRW1}) and (\ref{FRW2}) into Eq.(\ref{Inflaton}), we obtain
\begin{equation}
 \label{Inflaton2} \mathcal{A} \ddot{\phi} + 3 H \mathcal{B} \dot{\phi} + \tilde{V}_{\phi} = 0,
\end{equation}
where
\begin{eqnarray}
  \mathcal{A} & = & \mathcal{L}_X c_s^{- 2} + \frac{3 M_p^2 f_{\phi}^2}{2 f}, \\
  \mathcal{B} & = & \mathcal{L}_X + r + \frac{3 M_p^2 f_{\phi}^2}{2 f} -
  \frac{f_{\phi}}{2 f} \frac{\mathcal{L}_X \dot{\phi}}{3 H} + M_p^2
  \frac{f_{\phi}}{2 f} \frac{f_{\phi \phi} \dot{\phi}}{3 H},  \\
  \tilde{V}_{\phi} & = & V_{\phi} + \frac{2 f_{\phi}}{f} \mathcal{L} - \frac{f_{\phi}}{2 f} T s,
\end{eqnarray}
and $r=\frac{\Gamma}{3H}$ is the parameter that describes the damping strength of warm inflation. Now lets us define a dimensionless Hubble Slow-Roll (HSR) parameter $\epsilon_H$ as
\begin{eqnarray}
  \label{epsilon_H} \epsilon_H & = & - \frac{\dot{H}}{H^2} \nonumber\\
  & = & \frac{2 X \mathcal{L}_X + M_p^2 (f_{\phi}\ddot{\phi} + f_{\phi \phi} \dot{\phi}^2 - H f_{\phi} \dot{\phi}) + T s}{2 f M_p^2 H^2} \nonumber\\
  & = & \frac{X \mathcal{L}_X}{f
  M_p^2 H^2} + \frac{f_{\phi} \dot{\phi}}{2 f H} \frac{\ddot{\phi}}{H
  \dot{\phi}} + \frac{f_{\phi \phi} \dot{\phi}}{2 f_{\phi} H } \frac{f_{\phi}
  \dot{\phi}}{f H } - \frac{f_{\phi} \dot{\phi}}{2 f H} + \frac{T s}{2 f M_p^2
  H^2} \nonumber\\
  & = & \delta_X + \frac{1}{2} \delta_f \delta_{\phi} + \frac{1}{2} \delta_f
  \eta_f - \frac{1}{2} \delta_f + \frac{1}{2} \delta_T,
\end{eqnarray}
where
\begin{eqnarray}
\delta_X \equiv \frac{X \mathcal{L}_X}{f M_p^2 H^2}, \quad
\delta_{\phi} \equiv  \frac{\ddot{\phi}}{H \dot{\phi}} ,\quad
\delta_f \equiv \frac{f_{\phi} \dot{\phi}}{H f}  ,\quad
\eta_f  \equiv  \frac{f_{\phi \phi} \dot{\phi}}{H f_{\phi}},\quad
\delta_T  \equiv  \frac{T s}{ f M_p^2 H^2}.
\end{eqnarray}
Accordingly, inflation then takes place when condition $\epsilon_H <1 $ is satisfied, implying $\ddot{a}> 0$ and will terminate when $\epsilon_H = 1$.

The exact solutions of the dynamical equations governing inflation are often difficult to obtain. This motivates the use of slow-roll approximations, under which the Hubble parameter varies only slightly per e-fold, so that \(\dot{H} \ll H^2\), while the leading derivative terms are neglected: \(\ddot{\phi} \ll H\dot{\phi}\) and \(\dot{s} \ll Hs\). These approximations imply that the energy density is dominated by the potential, the inflaton evolves slowly, and radiation is produced quasistatically. The first condition, \(\dot{H} \ll H^2\), further suggests that each term in Eq. (\ref{epsilon_H}) should be small. This assumption should be applied with care, however, since the sum of the terms may also be small
\begin{equation}
\{ \delta_X, \delta_{\phi}, \delta_f, \eta_f, \delta_T \} \ll 1,
\end{equation}
which mean
\begin{equation}
X \mathcal{L}_X = X K_X \sim K \ll V, \quad \ddot{\phi} \ll H \dot{\phi}, \quad f_{\phi} \dot{\phi} \ll H f, \quad f_{\phi \phi} \dot{\phi} \ll H f_{\phi} \quad T s \ll V.
\end{equation}
As a result, the slow-roll equations take the following form
\begin{eqnarray}
  \label{SlowInflaton} 3 H \mathcal{B} \dot{\phi}  &\simeq & -\tilde{V}_{\phi}, \\
   \label{SlowFRW} 3 f M_p^2 H^2 &\simeq & V, \\
   \label{SlowEntropy} T s &\simeq & r \dot{\phi}^2,
\end{eqnarray}
and $\mathcal{B}$ and $\tilde{V}_{\phi}$  reduce to
\begin{eqnarray}
 \mathcal{B} & = & \mathcal{L}_X + r + M_p^2  \frac{3 f_{\phi}^2}{2 f} -
  \frac{\mathcal{L}_X}{6} \frac{f_{\phi} \dot{\phi}}{H f} + f M_p^2
  \frac{f_{\phi} \dot{\phi}}{6 H f} \frac{f_{\phi \phi}}{f} \nonumber \\
  & = & \mathcal{L}_X \left( 1 - \frac{\delta_f}{6} \right)  + r + 3 f
  \frac{M_p^2}{2}  \frac{f_{\phi}^2}{f^2} \left( 1 + \frac{\eta_f}{9} \right) \nonumber \\
  & \simeq & \mathcal{L}_X + r + 3 f \frac{M_p^2}{2} \frac{f_{\phi}^2}{f^2}, \\
\tilde{V}_{\phi}  & \simeq & V_{\phi} - 2 \frac{f_{\phi}}{f} V.
\end{eqnarray}
Compared with minimally coupled standard canonical inflation, the effective damping coefficient is enhanced by a factor of $(\mathcal B\simeq\mathcal L_X+r+3M_p^2f_\phi^2/(2f^2))$ owing to the combined contributions from the noncanonical kinetic term, thermal dissipation, and nonminimal coupling, which indicates that NMSC noncannonical warm inflation can exhibit a strongly overdamped evolution. Meanwhile, the effective potential gradient is modified to $(\tilde V_\phi\simeq V_\phi-2(f_\phi/f)V = V_\phi[1-2(f_\phi/f)(V/V_\phi)] )$, which can reduce the driving force of the inflaton when $|2(f_\phi/f)(V/V_\phi)|<1$. For example, when $V(\phi) = \lambda \phi^4$, \ $f (\phi) = 1 + \xi \phi^2 / M_p^2$, $|2(f_\phi/f)(V/V_\phi)|$ is smaller than 1.  Thus, the combined effects of enhanced damping and a potentially flattened effective potential facilitate slow-roll evolution, allowing inflation to occur even for relatively steep potentials.

Furthermore, the number of e-folding in NMSC noncannonical warm inflationary scenario is defined as follows
\begin{equation}
  N = \int H d t = \int \frac{H}{\dot{\phi}} d \phi \simeq - \frac{1}{M_p^2} \int_{\phi_{\ast}}^{\phi_{\text{end}}} \frac{\mathcal{B}}{f} \frac{V}{ \tilde{V}_{\phi}}  d \phi,
\end{equation}
where $\phi_{\ast}$ denotes the value of the inflaton field at the Hubble crossing time and $\phi_{\text{end}}$ represents the value of the inflaton field at the end of inflation. The e-folding number can be enhanced by the combined effects of the increased effective damping coefficient and the modified potential gradient, allowing NMSC noncannonical warm inflation to achieve sufficient inflationary expansion with a relatively small inflaton field excursion.

\section{Stability analysis}
\label{sec:slowroll}

In order to make a systematic stability analysis, we define the customary dimensionless Potential Slow-roll (PSR) parameters as
\begin{equation}
  \epsilon = \frac{M^2_{p }}{2} \left( \frac{V_{\phi}}{V} \right)^2, \quad
  \eta =M^2_{p } \frac{V_{\phi \phi}}{V}, \quad
  \beta = M^2_{p } \frac{V_{\phi}\Gamma_{\phi}}{V_{} \Gamma},
\end{equation}
which are supplemented with two more parameters about the temperature dependence, namely
\begin{equation}
  b = \frac{T V_{\phi T}}{V_{\phi}}, \quad  c = \frac{T \Gamma_T}{\Gamma}.
\end{equation}

Compared with conventional warm inflation, the slow-roll stability of the present model requires not only that perturbations in the dissipative and radiation sectors do not destabilize the slow-roll attractor, but also that the effective gravitational coupling varies sufficiently slowly during inflation. Since this variation is governed by the nonminimal coupling function $f(\phi)$, its first and second derivatives must be properly constrained. To characterize these additional effects, we introduce the following slow-roll parameters associated with the nonminimal coupling function:
\begin{equation}
  \varphi = \frac{M^2_{p }}{2} \left( \frac{f_{\phi}}{f}
  \right)^2, \quad \kappa = M^2_{p } \frac{f_{\phi \phi}}{f}, \quad \gamma =
  M^2_{p } \frac{V_{\phi} f_{\phi}}{V f}.
\end{equation}
Using $\tilde{V}_{\phi} = V_{\phi} - 2 \frac{ f_{\phi}}{f} V = V_{\phi} \left( 1 - \frac{\gamma}{\epsilon} \right) \equiv V_{\phi} \mathcal{Z}$, we can redefine the parameters related to the potential slope $V_{\phi}$ as
\begin{equation}
  \tilde{\epsilon} = \frac{M^2_{p }}{2} \left( \frac{\tilde{V}_{\phi}}{V} \right)^2, \quad
  \tilde{\eta} = M^2_{p } \frac{\tilde{V}_{\phi \phi}}{V}, \quad
  \tilde{\beta} = M^2_{p } \frac{\tilde{V}_{\phi} \Gamma_{\phi}}{V_{} \Gamma},\quad
  \tilde{b} = \frac{T V_{\phi T}}{\tilde{V}_{\phi}}, \quad
  \tilde{\gamma} =M^2_{p } \frac{\tilde{V}_{\phi} f_{\phi}}{V_{} f}.
\end{equation}
And their relations are
\begin{equation}
\tilde{\epsilon} = \epsilon  \mathcal{Z}^2, \quad
\tilde{\eta} = \eta - 2 \gamma - 2 \kappa + 4 \varphi,\quad
\tilde{\beta} = \beta  \mathcal{Z},\quad
\tilde{b} = \frac{b}{\mathcal{Z}},\quad
\tilde{\gamma} = \gamma \mathcal{Z}.
\end{equation}

Inflation is commonly described using the slow-roll approximation, in which the highest-order terms in Eqs.(\ref{entropy}) and (\ref{Inflaton2}) are neglected. The resulting conditions imply that the inflaton¡¯s energy is potential-dominated, the field evolves gradually, and radiation is produced quasi-statically. However, applying these approximations to obtain the slow-roll equations, Eqs.(\ref{SlowInflaton})-(\ref{SlowEntropy}), raises an important question: under what conditions do these slow-roll equations can portray the system well? To address this, we define \(u=\dot{\phi}\) and rewrite Eqs.(\ref{entropy}) and (\ref{Inflaton2}) as follows:
\begin{equation}
 \label{dotu} \dot{u} = - \mathcal{A}^{-1} (3 H \mathcal{B} u + \tilde{V}_{\phi}) = 0,
\end{equation}
\begin{equation}
  \label{dots} \dot{s} = - 3 H s + \frac{\Gamma u^2}{T}.
\end{equation}

The variables \(u_0\) and \(s_0\) denote slow-roll solutions satisfying the equations below
\begin{eqnarray}
 \label{SlowPhi} u_0 & = & - \frac{\tilde{V}_{\phi}}{3 H_0 \mathcal{B}}, \\
  \label{SlowS} s_0 & = & r u_0^2 T^{- 1}.
\end{eqnarray}
The relation \(\dot{\phi}_0=u_0\) is introduced simply as a convenient definition for the stability analysis as follows. Assessing the validity of the slow-roll approximation requires a linear stability analysis. This involves perturbing the exact solutions about the slow-roll solutions and determining the conditions under which the system remains close to them over many Hubble times. In other words, the slow-roll solutions must act as attractors of the dynamical system. We therefore perturb the variables \(\phi\), \(u\), and \(s\) around their slow-roll solutions as follows:
\begin{equation}
  \phi = \phi_0 + \delta \phi, \quad u = u_0 + \delta u, \quad s = s_0 + \delta s.
\end{equation}
The perturbations \(\delta\phi\), \(\delta u\), and \(\delta s\) are assumed to be much smaller than their corresponding background quantities \(\phi_0\), \(u_0\), and \(s_0\). We analyze stability around the slow-roll solutions to determine the conditions under which they can serve as genuine attractors of the dynamical system.

Using the new variable $u$, $X = \frac{1}{2} u^2$, then $\delta X = u \delta u$, $\delta \mathcal{L}_X = \mathcal{L}_{X X} u \delta u$. Varying the Friedmann equation (\ref{FRW1}) we obtain
\begin{eqnarray}
  (2 f H + f_{\phi} u) \delta H  & = &
  \frac{1}{3 M_p^2} (\mathcal{L}_X c^{- 2}_s u \delta u + V_{\phi} \delta \phi
  + T \delta s) \nonumber \\
  &&- (f_{\phi} H^2 \delta \phi + H f_{\phi \phi} u
  \delta \phi + H f_{\phi} \delta u).
\end{eqnarray}
And through the thermal relation (\ref{thermal}), we have
\begin{equation}
  \delta s = - V_{T T} \delta T - V_{\phi T} \delta \phi.
\end{equation}
Thus
\begin{equation}
  \delta T = - \frac{1}{V_{T T}} (\delta s + V_{\phi T} \delta \phi).
\end{equation}
We have the formula of $T s_T \simeq 3 s$ when the thermal correction to the potential is small, and using the formula we get
\begin{equation}
  \delta T = \frac{1}{3 s} (T \delta s + V_{\phi} b \delta \phi).
\end{equation}
By applying the definitions of the slow-roll parameters and varying Eqs.(\ref{dotu}) and (\ref{dots}), we can get
\begin{eqnarray}
\left(\begin{array}{c}
     \delta \dot{\phi}\\
     \delta \dot{u}\\
     \delta \dot{s}
   \end{array}\right) = E \cdot \left(\begin{array}{c}
     \delta \phi\\
     \delta u\\
     \delta s
   \end{array}\right) - F.
\end{eqnarray}
The matrices $E$ and $F$ can be expressed as
\begin{equation}
  E = \text{} \left( \begin{array}{c}
    0\\
    A\\
    C
  \end{array} \begin{array}{c}
    1\\
    \lambda_1\\
    D
  \end{array} \begin{array}{c}
    0\\
    B\\
    \lambda_2
  \end{array} \right) ,  \qquad
  F = \left( \begin{array}{c}
    0\\
    \dot{u}_0\\
    \dot{s}_0
  \end{array} \right).
\end{equation}
The elements of matrix $E$ are derived through calculations:
\begin{eqnarray}
  A & = & 3 H^2 \mathcal{A}^{- 1} \left[ \frac{(\mathcal{L}_X +M)}{\mathcal{B}} f \tilde{\epsilon} - f \tilde{\eta} + \frac{r}{\mathcal{B}}
  f \tilde{\beta} - \frac{\mathcal{B}^2 }{r} \tilde{b}^2 + \mathcal{B} \tilde{b} c + \frac{(\mathcal{L}_X + M)}{2 \mathcal{B}} f \tilde{\gamma} \right], \\
  B & = & \frac{H T}{u \mathcal{} } \mathcal{A}^{- 1} \left[ -
  \frac{(\mathcal{L}_X + M)}{\mathcal{B}^2 } f \tilde{\epsilon} - c +
  \frac{\mathcal{B}}{r} \tilde{b} - \frac{2}{3} \frac{f}{\mathcal{B}}
  \tilde{\gamma} \right], \\
  C & = & \frac{3 H^2 u}{T} \left[ \frac{r}{\mathcal{B}} f \tilde{\epsilon} -
  \frac{r}{\mathcal{B}} f \tilde{\beta} + \mathcal{B} (1 - c) \tilde{b} +
  \frac{r}{2 \mathcal{B}} f \tilde{\gamma} \right], \\
  D & = & \frac{H u}{T} \left[ 6 r - \frac{r \mathcal{L}_X c^{-
  2}_s}{\mathcal{B}^2 } f \tilde{\epsilon} - \frac{3 r }{2
  \mathcal{B}} f \tilde{\gamma} \right], \\
  \lambda_1 & = & - 3 H \left( 1 + \frac{r}{\mathcal{A} } \right) -
  \mathcal{L}_X c^{- 2}_s H \mathcal{A}^{- 1} \left[ \frac{(\mathcal{L}_X +
  M)}{\mathcal{B}^2} f \tilde{\epsilon} + \frac{f}{2 \mathcal{B}}
  \tilde{\gamma} \right], \\
  \lambda_2 & = & - H \left[ (4 - c) + \frac{r}{\mathcal{B}^2} f
  \tilde{\epsilon} \right],
\end{eqnarray}
where $M= 3 f \frac{M_p^2}{2} \frac{f_{\phi}^2}{f^2}=3f \varphi$. In the minimal coupling limit $f(\phi)=1$, all the matrix elements consistently reduce to those of noncanonical warm inflation obtained in Ref.\cite{ZhangZhu2014}.

Whether a warm-inflation dynamical system approaches its attractor depends on the eigenvalues of the matrix \(E\). Specifically, the slow-roll solutions can act as attractors when the eigenvalues are negative or, possibly, positive but of order $\mathcal{O} \left( f \frac{\tilde{\epsilon}}{\mathcal{B}} \right)$, corresponding to slow growth, and when the forcing term \(F\) is sufficiently small \cite{Moss:2008yb}. The latter requires \(\left|\dot{u}_0/(H u_0)\right|\ll 1\) and \(\left|\dot{s}_0/(H s_0)\right|\ll 1\). We first examine \(F\) by differentiating the slow-roll equations (\ref{SlowPhi})and (\ref{SlowS}) with respect to time and using the following relations:
\begin{eqnarray}
  \label{dotGamma} \dot{\Gamma}  = - 3 H^2 \left( \frac{r}{\mathcal{B}} f \tilde{\beta} +\mathcal{B} \tilde{b} c - \frac{\dot{s}}{3 H s} r c \right),
\end{eqnarray}
\begin{eqnarray}
  \label{dotr} \dot{r} =  - H \left[ \left( \frac{r}{\mathcal{B}} f \tilde{\beta} + \mathcal{B}
  \tilde{b} c - \frac{\dot{s}}{3 H s} r c \right) - \frac{r }{\mathcal{B}} f \left(
  \tilde{\epsilon} + \frac{1}{2} \tilde{\gamma} \right) + \frac{r T \dot{s}}{6
  H V} \right],
\end{eqnarray}
we can arrive at the leading order of slow-roll parameters
\begin{eqnarray}
  \frac{\dot{u}_0}{H u_0} & =&  \frac{1}{\Delta} \left\{ \left[ (4 - c)\frac{\mathcal{B}}{\mathcal{A}} - 4 \frac{r}{\mathcal{A}}  \right]
  \frac{f}{\mathcal{B}} \left( \tilde{\epsilon} + \frac{1}{2} \tilde{\gamma}
  \right) + 4 \frac{r}{\mathcal{A}} \frac{f}{\mathcal{B}} \tilde{\beta}\right.  \nonumber \\
 && \left. +  (c -4) \frac{f}{\mathcal{A}} \tilde{\eta} - 3 \frac{\mathcal{B}^2}{\mathcal{A} r} \tilde{b}^2 + 3 c \mathcal{} \frac{\mathcal{B}}{\mathcal{A} r} \tilde{b} \right\},
\end{eqnarray}
\begin{eqnarray}
  \frac{\dot{s}_0}{H s_0}& = & \frac{1}{\Delta} \left\{ 3 \left( 1 + 2
  \mathcal{\frac{B}{A}} - \frac{r}{\mathcal{A}} \right)
  \frac{f}{\mathcal{B}} \left( \tilde{\epsilon} + \frac{1}{2} \tilde{\gamma}
  \right) - 3 \left( 1 - \frac{r}{\mathcal{A}} \right) \frac{f}{\mathcal{B}}
  \tilde{\beta} \right.  \nonumber \\
  && \left. - 6 \frac{f}{\mathcal{A}} \tilde{\eta} - 6
  \frac{\mathcal{B}^2}{\mathcal{A} r} \tilde{b}^2 + 3 \left[ (1 + c)
  \frac{r}{\mathcal{A}} + (1 - c) \mathcal{} \right] \frac{\mathcal{B}}{r}
  \tilde{b} \right\},
\end{eqnarray}
with $\Delta$ being defined as
\begin{eqnarray}
  \Delta = (4 - c) + (c + 4) \frac{r}{\mathcal{A}} - \frac{2\mathcal{B}}{\mathcal{A}} \tilde{b} - \frac{4 r}{3 \mathcal{A}} \frac{f}{\mathcal{B}} \tilde{\gamma}.
\end{eqnarray}
The Hubble slow-roll parameter $\epsilon_H$ should also vary slowly, i.e. $\epsilon_H = -\frac{\dot{H}}{H^2} \simeq   \frac{f}{\mathcal{B}} \left(\tilde{\epsilon} + \frac{1}{2} \tilde{\gamma} \right) \ll 1$ which yields sufficient conditions for meeting the requirements above. Based on $\left| \frac{\dot{u}}{H u} \right| \ll 1, \left| \frac{\dot{s}}{H s} \right| \ll 1$, we can obtain
\begin{equation}
  \tilde{\epsilon} \ll \frac{\mathcal{B}}{f}, \quad \tilde{\beta} \ll
  \frac{\mathcal{B}}{f}, \quad \tilde{\eta} \ll \frac{\mathcal{A}}{f}, \quad
  \tilde{b} \ll \frac{ \text{min} \left\{ \mathcal{A},r \right\} }{\mathcal{B}},  \quad \tilde{\gamma} \ll \frac{\mathcal{B}}{f}.
\end{equation}
 In the minimal coupling limit $f(\phi)\to 1$, all coupling-dependent corrections vanish, yielding $\mathcal{A}\to\mathcal{L}_X c_s^{-2}$, $\mathcal{B}\to\mathcal{L}_X+r$ and $\tilde{\epsilon}\to\epsilon$, $\tilde{\beta}\to\beta$, $\tilde{\eta}\to\eta$, $\tilde{b}\to b$. Consequently, the standard slow-roll conditions of noncanonical warm inflation are recovered\cite{ZhangZhu2014}. Compared with noncanonical warm inflation with minimal coupling, the constraints derived in the present model exhibit additional flexibility arising from the nonminimal scalar-curvature coupling. In particular, the effective potential gradient is modified according to $\widetilde V_\phi=V_\phi(1-\gamma/\epsilon)\equiv  V_\phi \mathcal Z$, which leads to a suppression of the effective driving force when $0<|\mathcal Z|<1$. Consequently, the slow-roll conditions expressed in terms of the original potential parameters can be relaxed in suitable regions of parameter space, allowing inflation to occur even for relatively steep potentials that would otherwise be strongly constrained in conventional noncanonical warm inflation. Furthermore, the nonminimal coupling modifies the dynamical coefficients through $\mathcal A=\mathcal L_Xc_s^{-2}+3f\varphi$ and $\mathcal B \simeq \mathcal L_X+ r +3f\varphi$. These modifications provide additional freedom to satisfy the slow-roll conditions through the combined effects of noncanonical kinetic terms, thermal dissipation, and nonminimal coupling. The slow-roll condition for $b$ implies that the thermal correction to the inflationary potential should remain small, as is the case for warm inflation in GR \cite{Moss:2008yb,delCampo:2010by}. Thus, the total energy density can take a nearly separable form $\rho \simeq \rho(\phi,X)+\rho_r - 3 M_p^2 H f_{\phi} \dot{\phi}$.

So far, we have derived conditions on all model parameters except \(c\). To constrain \(c\), we now calculate the eigenvalues of the matrix \(E\). Using the slow-roll conditions obtained above, we find that
\begin{eqnarray}
  \det (\lambda I - E) & = & \left| \text{} \begin{array}{c}
    \lambda\\
    - A\\
    - C
  \end{array} \begin{array}{c}
    - 1\\
    \lambda - \lambda_1\\
    - D
  \end{array} \begin{array}{c}
    0\\
    - B\\
    \lambda - \lambda_2
  \end{array} \right| \nonumber\\
  & = & (\lambda - \lambda_2) [\lambda (\lambda - \lambda_1) - A] - B D
  \lambda - B C \nonumber\\
  & = & 0
\end{eqnarray}
has a very small eigenvalue $\lambda \simeq \frac{B C - A \lambda_2}{\lambda_1 \lambda_2 - B D - A} \ll \lambda_1, \lambda_2.$ The other two eigenvalues satisfy $\lambda^2 - (\lambda_1 + \lambda_2) \lambda + \lambda_1 \lambda_2 - B D = 0 $. The two eigenvalues are both negative when $\lambda_1 + \lambda_2 < 0$ and $\lambda_1 \lambda_2- B D > 0$. Based on the above element of matrix $E$, we get
\begin{eqnarray}
  \lambda_1 + \lambda_2 \simeq  - H \left[ 3 \left( 1 + \frac{r}{\mathcal{A} } \right) + (4 - c) \right],
\end{eqnarray}
and
\begin{eqnarray}
  \lambda_1 \lambda_2 - B D \simeq 3H^2  \left[ (4 - c) + (c + 4) \frac{r}{\mathcal{A}}\right].
\end{eqnarray}
Using the above conditions for satisfying the slow-roll requirements, when $\lambda_1 + \lambda_2 < 0$ and $\lambda_1 \lambda_2- B D > 0$, we can obtain
\begin{equation}
\label{c}| c | < 4,
\end{equation}
which implies that the temperature dependence of the dissipative coefficient must lie within the range $\Gamma \propto (T^{-4},T^4)$. Moreover, Eqs.(\ref{SlowInflaton})-(\ref{SlowEntropy}) yield $\frac{\rho_r}{V}\simeq \frac{r f}{2\mathcal{B}^2}\,\tilde{\epsilon}\ll 1,$
indicating that radiation remains subdominant during slow roll. This is consistent with the assumption that inflation is dominated by the potential energy.

\section{Cosmological perturbations}
\label{sec:perturbations}
We now develop the theory of cosmological perturbations in the NMSC noncanical warm inflationary model. Since we are working in the context of a cosmological system, metric perturbations as well as field and thermal fluctuation should be included. A defining feature of dissipative inflation, in contrast to cold inflation, is that density fluctuations arise from thermal fluctuations in the radiation field rather than from quantum fluctuations. Through the damping terms in the inflationary dynamical equations, these thermal fluctuations couple to the inflaton, and their amplitude is determined by the fluctuation-dissipation theorem. Consequently, both entropy and curvature perturbations contribute to density fluctuations in warm inflation. During inflation, radiation is subdominant, so its thermal fluctuations contribute primarily to entropy perturbations. In a system with a heat bath, entropy perturbations decay on superhorizon scales \cite{Hall:2003zp,Moss:2008yb}; consequently, one should keep track of curvature perturbations.

Primordial cosmological perturbations are commonly described by the curvature perturbation \(\mathcal{R}\) on uniform-density hypersurfaces. This quantity is useful because, in simple models, it is conserved on large scales even beyond linear order. At linear order in single-field slow-roll inflation, the curvature perturbation on uniform-density hypersurfaces is given by the gauge-invariant combination $\mathcal{R}=\Psi+\frac{H}{\dot{\rho}}\delta\rho,$ where \(\Psi\) denotes the spatial metric perturbation and \(\delta\phi\) the perturbation of the homogeneous inflaton field. For convenience, we work in the spatially flat gauge, and accordingly, the perturbation is given by $\mathcal{R}=\frac{H}{\dot{\rho}}\delta\rho$ which in the slow-roll regime becomes $\mathcal{R}=\frac{H}{\dot{\phi}}\delta\phi$.

For small perturbations, we expand the inflaton field as $\Phi(x,t)=\phi(t)+\delta\phi(x,t),$ where \(\phi(t)\) is the homogeneous background and \(\delta\phi(x,t)\ll\phi(t)\) is its perturbation. The perturbation is the linear response of the thermal system to the stochastic noise \(\xi\). In the high-temperature limit, \(T\to\infty\), this noise is Markovian, with correlation function $\left\langle\xi(\mathbf{k},t)\xi(-\mathbf{k}',t')\right\rangle=2\Gamma T a^{-3}(2\pi)^3\delta^3(\mathbf{k}-\mathbf{k}')\delta(t-t')$ \cite{Hall:2003zp,Gleiser:1993ea}. Including the noise term and substituting the inflaton-field expansion gives the following second-order Langevin equation in Fourier space:
\begin{equation}
  \label{LangevinEq} \mathcal{A} \delta \ddot{\phi}_\mathbf{k} + 3 H \mathcal{D} \delta \dot{\phi}_\mathbf{k} + \left( \mathcal{L}_X \frac{k^2}{a^2} + V_{\phi\phi}+ \mathcal{S} \right) \delta \phi_\mathbf{k} = \xi_\mathbf{k},
\end{equation}
where
\begin{eqnarray}
 \mathcal{A} & = &   \mathcal{L}_X c_s^{-2} + 3f \varphi ,\\
 \mathcal{D} & = & \left[ \mathcal{L}_X \left( 1 + \frac{\delta_f}{3} \right) +
  \mathcal{L}_{X X} \dot{\phi}^2 \left( 1 - \frac{\delta_f}{6} \right) + 3 f
  \varphi \left( 1 + \frac{2 \eta_f}{9} \right) + r \right],\\
  \mathcal{S}  & = & 3 H^2 \left[ 3 f \varphi (2 \eta_f - \delta_f) + \frac{1}{3} f
  (\kappa - 2 \varphi) \left( \frac{2 \mathcal{L}}{f M_p^2 H^2} - \frac{1}{2} \delta_T -
  \delta_X + \frac{1}{2} \delta_f \eta_f \right)  \right. \nonumber \\
  && \left.  + r \delta_{\Gamma} - 2 \delta_f \frac{V_{\phi}}{3 H \dot{\phi}} \right],
\end{eqnarray}
and $\delta_\Gamma \equiv \frac{\Gamma_\phi \dot{\phi}}{H \Gamma}$. The second-order Langevin equation (\ref{LangevinEq}) describes the stochastic evolution of inflaton fluctuations in the NMSC noncanonical warm inflationary scenario. $\mathcal A$ and $\mathcal D$ represent the effective inertia and damping coefficients of inflaton fluctuations, respectively, incorporating modifications arising from the noncanonical kinetic structure and nonminimal coupling. The gradient term $\mathcal L_X k^2/a^2$ governs the spatial propagation of perturbations, while $V_{\phi\phi}+\mathcal S$ determines their effective mass squared, with $\mathcal S$ encoding additional corrections associated with the nonminimal coupling and dissipation.

To calculate the power spectrum at Hubble crossing, we use the fact that this occurs well within the slow-roll regime. As established by the stability analysis, the slow-roll solution remains valid there, allowing us to neglect the inertial terms and reduce the Langevin equation to first order:
\begin{equation}
 \label{Langevin} 3 H \mathcal{D}\delta \dot{\phi}_\mathbf{k} + \left( \mathcal{L}_X \frac{k^2}{a^2} + \mathcal{M}^2 \right) \delta \phi_\mathbf{k} = \xi_\mathbf{k},
\end{equation}
where the mass term $\mathcal{M}^2 \equiv V_{\phi \phi} + \mathcal S$. And $\mathcal{D}$ and $\mathcal{M}^2$ during the slow-roll regime reduce to
\begin{eqnarray}
  \mathcal{D}& \simeq &  \mathcal{A} + r, \\
  \mathcal{M}^2 & \simeq &  3 H^2 (f \tilde{\eta} + r\frac{f}{\mathcal B} \tilde{\beta}).
\end{eqnarray}
The approximate analytic solution of Eq.(\ref{Langevin}) is
\begin{eqnarray}
\label{LangevinSolution} \delta \phi_\mathbf{k}(t) & \approx &  \frac{1}{3 H \mathcal{D}}\exp \left[ - \frac{t - t_0}{\tau (\phi_0)} \right]\int_{t_0}^{t } \exp \left[ \frac{t^{\prime} - t_0}{\tau (\phi_0)}  \right] \xi  (\mathbf{k}, t^{\prime}) d t^{\prime} \nonumber\\
      &&+ \delta \phi (\mathbf{k}, t_0)  \exp \left[ - \frac{t - t_0}{\tau (\phi_0)}\right],
\end{eqnarray}
where \(\tau = \frac{3 H  \mathcal{D}}{\mathcal{L}_X k^2/a^2 +\mathcal{M}^2}\), characterizes the efficiency of thermalization. The first term on the right-hand side drives \(\delta\phi\) toward thermal equilibrium, while the final term retains the dependence on its initial value. This memory of the initial perturbation becomes negligible as time passes.

Because the thermal effects in Eq.(\ref{Langevin}) depend on physical wave numbers, \(k_p=k/a\), where \(k\) is the comoving wave number. In an expanding universe, Eq.(\ref{LangevinSolution}) shows that modes with larger \(k_p^2\) relax more rapidly. Once the physical wave number of a \(\delta\phi(x,t)\) mode falls below the freeze-out wave number \(k_F\), the mode is effectively unaffected by the thermal noise \(\xi(\mathbf{k},t)\) over a Hubble time \cite{Berera:1999ws}. Accordingly, \(k_F\) is defined by $\frac{\mathcal{L}_X k_F^2+\mathcal{M}^2} {3H^2\mathcal{D}}=1.$ During slow-roll inflation, the mass term can be negligible because $\frac{\mathcal{M}^2}{3H^2 \mathcal{D}} = \frac{f}{\mathcal{A}+r} \tilde{\eta} + \frac{r}{\mathcal{A}+r} \frac{f}{\mathcal{B}} \tilde{\beta} \ll 1$. Thus $k_F$ is given by
\begin{equation}
k_F  = \sqrt{\frac{3 H^2 \mathcal{D}}{\mathcal{L}_X}}.
\end{equation}
Compared with conventional canonical and noncanonical warm inflation in general relativity, the present model introduces an additional nonminimal coupling contribution to the effective damping coefficient, thereby enhancing the freeze-out scale of inflaton fluctuations for a positive coupling contribution. This means that a mode ceases to efficiently interact with thermal noise at a larger physical wavenumber, corresponding to an earlier freeze-out. Since thermal fluctuations generally satisfy $\langle\delta\phi^2\rangle\propto k_F T$ a larger $k_F$ enhances the inflaton fluctuation amplitude and the scalar power spectrum, thereby typically further suppressing the tensor-to-scalar ratio.

Based on the field perturbation relation $\delta \phi^2 = \frac{k_F  T}{2 \pi^2}$ in warm inflation \cite{Berera:1999ws,Taylor:2000ze}, and using $P_\mathcal{R} = \left( \frac{H}{\dot{\phi}}\right)^2 \delta \phi^2$, we can finally get the scalar power spectrum in NMSC noncanonical warm inflationary model:
\begin{eqnarray}
    \label{scalarspectrum} P_{\mathcal{R}}=\frac{H^3 T}{2 \pi^2 \dot{\phi}^2} \sqrt{\frac{3 \mathcal{D}}{\mathcal{L}_X}}.
\end{eqnarray}
Also, the spectral index is given by
\begin{eqnarray}
  \label{spectralindex} n_s - 1 & = & \frac{d \ln P_R}{d \ln k} \nonumber \\
  & \simeq & 3 \frac{\dot{H}}{H^2} + \frac{\dot{T}}{H T} - 2
  \frac{\ddot{\phi}}{H \dot{\phi}} + \frac{\dot{\mathcal{D}}}{2 H \mathcal{D}}
  - \frac{\dot{\mathcal{L}}_X}{2 H \mathcal{L}_X}\nonumber \\
  & = & - 3 \epsilon_H + \theta_T - 2 \delta_{\phi} + \frac{1}{2}
  \delta_{\mathcal{D}} - \frac{1}{2} \delta_{\mathcal{L}_X},
\end{eqnarray}
where
\begin{eqnarray}
  \theta_T & = &  \frac{\dot{s}}{3 H s} - \frac{\mathcal{B}}{r} \tilde{b},   \\
  \delta_{\mathcal{D}} & = & \frac{r}{\mathcal{L}_X c_s^{- 2} + 3 f \varphi + r} \left( -
  \frac{f}{\mathcal{B}} \tilde{\beta} - \frac{\mathcal{B} }{r} \tilde{b} c +
  \frac{\dot{s}}{3 H s} c + \epsilon_H \right) \nonumber \\
    && + \frac{\mathcal{L}_X (c_s^{-
  2} - 1)}{\mathcal{L}_X c_s^{- 2} + 3 f \varphi + r} (3 + 2 \kappa_X)
  \delta_{\phi} + \frac{3 f \varphi}{\mathcal{L}_X c_s^{- 2} + 3 f \varphi +
  r} (2 \eta_f - \delta_f), \\
  \delta_{\mathcal{L}_X} & = & (c_s^{- 2} - 1) \delta_{\phi},
\end{eqnarray}
and $\kappa_X \equiv \frac{X \mathcal L_{XX}}{ \mathcal L_{XXX}}$. Since all the parameters contributing to the scalar spectral index in Eq.(\ref{spectralindex}) are slow-roll suppressed,  the scalar spectral index $n_s$ remains close to unity.  Thus, the resulting power spectrum in our NMSC noncanonical warm inflationary model is nearly scale-invariant, in agreement with observations.

Tensor perturbations do not couple to the thermal bath. Consequently, as in standard inflation in the Jordan frame, gravitational waves arise solely from quantum fluctuations \cite{Diaz:2023tma}:
\begin{equation}
  P_\mathcal{T} = \frac{2}{f \pi^2} \left( \frac{H}{M_p} \right)^2.
\end{equation}
The spectral index of tensor perturbation is
\begin{equation}
  n_\mathcal{T} \simeq - 2 \epsilon_H + \frac{f}{\mathcal{B}}\tilde{\gamma} \simeq - 2\frac{f}{\mathcal{B}} \tilde{\epsilon},
\end{equation}
and the tensor-to-scalar ratio is
\begin{eqnarray}
  R = \frac{P_\mathcal{T}}{P_\mathcal{R}} = \frac{H}{T} \frac{8 f \tilde{\epsilon}}{\mathcal{B}^2} \sqrt{\frac{\mathcal{L}_X}{3 \mathcal{D}}}.
\end{eqnarray}
Also, the consistency relation becomes
\begin{eqnarray}
R = - \frac{H}{T} \frac{4 \sqrt{\mathcal{L}_X}}{\mathcal{B} \sqrt{3\mathcal{D}}} n_\mathcal{T},
\end{eqnarray}
which is not a fixed relation as in standard cold inflation anymore. Considering the slow-roll condition $\tilde{\epsilon} <\frac{\mathcal{B}}{f}$, we find that the energy scale of inflation merely bounds from above as
\begin{equation}
  R < \frac{H}{T} \frac{8 \sqrt{\mathcal{L}_X}}{\mathcal{B} \sqrt{3 \mathcal{D}}}.
\end{equation}
As shown in above result, thermal effects suppress the tensor-to-scalar ratio through the factor $H/T<1$, while nonminimal coupling can further reduce its upper bound by modifying the effective coefficients $\mathcal B$ and $\mathcal D$. Thus, the combined effects of nonminimal coupling, noncanonical dynamics, and thermal dissipation provide an additional mechanism for suppressing primordial tensor perturbations relative to scalar perturbations, facilitating a small tensor-to-scalar ratio consistent with observational constraints. These results suggest that a small \(R\) is favored in the NMSC noncanonical warm-inflation scenario.

The radiation energy density and universal temperature has the Stefan-Boltzmann relationship $\rho_r = \frac{\pi^2 g_{\star}}{30}T^4$ and using the slow-roll equations we obtained
\begin{eqnarray}
  \label{T/H}   \frac{T}{H} & = & \left( \frac{r}{g_{\ast} P_{\mathcal R}} \right)^{1 / 3} \left(\frac{45}{4 \pi^4} \right)^{1 / 3} \left( \frac{3
  \mathcal{D}}{\mathcal{L}_X} \right)^{1 / 6},
\end{eqnarray}
 the ratio \(T/H\) is slightly higher than in the GR limit of noncanonical warm inflation \cite{Moss:2008yb,ZhangZhu2014}. This increase makes thermal fluctuations more significant relative to Hubble expansion and helps maintain the warm-inflation condition \(T>H\). As Eq.(\ref{T/H}) shows, this condition is readily satisfied when \(r>g_*P_{\mathcal{R}}\). Given that \(g_*\sim\mathcal{O}(10^2)\) and \(P_{\mathcal{R}}\sim\mathcal{O}(10^{-9})\), even a small amount of dissipation can support warm inflation. Thus, warm inflation may provide a more realistic description of the very early Universe, including in the weak-dissipation regime. Furthermore, Eqs. (\ref{scalarspectrum}) and (\ref{T/H}) show that thermal fluctuations dominate over their quantum counterparts.

One evaluates the variation of the inflaton field for observable scales with $\Delta N \simeq 4.6 $ corresponding to $1 \leq l \leq 100$ as follows
\begin{eqnarray}
\frac{\Delta\phi}{M_{P}} = \frac{\dot\phi \Delta N }{ M_{P} H} \simeq 3.03 \left( \frac{T}{H} \right)^{1/2} \left(  \frac{
  \mathcal{D}}{\mathcal{L}_X}  \right)^{1/4} \left( f R\right)^{1/2}
\end{eqnarray}
The standard Lyth relation is modified in the present NMSC noncanonical warm inflationary scenario, where the inflaton field excursion is jointly governed by the tensor-to-scalar ratio, nonminimal coupling, noncanonical kinetic effects, and thermal dissipation. This modified relation provides additional flexibility in regulating the field excursion, allowing a suppressed tensor-to-scalar ratio to coexist with different field excursion scales. In particular, sub-Planckian field excursions can be realized in suitable parameter regions, potentially alleviating the large-field excursion problem encountered in conventional inflationary models.

\section{Conclusions and discussions}
\label{sec:conclusion}
In this work, we have systematically investigated the background dynamics, slow-roll stability, and primordial cosmological perturbations of a noncanonical warm inflationary model with a nonminimal coupling between the inflaton field and scalar curvature. By considering a general noncanonical Lagrangian and the coupling function $f(\phi)$, we have derived the modified inflationary equations in the Jordan frame. Our analysis reveals that the nonminimal coupling modifies both the effective damping coefficient and the potential gradient governing the inflaton evolution. In particular, the combined contributions from noncanonical kinetic effects, thermal dissipation, and nonminimal coupling enhance the effective damping, while the modified potential gradient can be reduced in suitable regions of parameter space. These effects facilitate slow-roll inflation and may allow sufficient accelerated expansion even for relatively steep potentials and smaller inflaton field excursions.

We have further performed a linear stability analysis of the slow-roll solutions by examining perturbations around the inflationary background. The stability conditions are expressed in terms of generalized slow-roll parameters that explicitly incorporate the effects of nonminimal coupling. We find that the additional coupling contributions modify the constraints on the potential slow-roll parameters and provide greater flexibility in realizing a stable slow-roll attractor. In particular, the stability requirement on the temperature dependence of the dissipation coefficient leads to the familiar sufficient condition $|c|<4$. In the minimal-coupling limit, our results consistently reduce to those of conventional noncanonical warm inflation. Thus, the nonminimal coupling extends the allowed slow-roll dynamics without altering the essential stability mechanism of warm inflation.

To investigate primordial fluctuations, we have employed the stochastic approach and derived the second-order Langevin equation for inflaton perturbations. The nonminimal coupling introduces additional corrections to the effective inertia, damping, and mass terms of the perturbation equation. Under the slow-roll approximation, the Langevin equation reduces to an overdamped stochastic equation, from which the freeze-out scale and scalar curvature power spectrum can be obtained. We find that the enhanced effective damping can increase the freeze-out wavenumber and strengthen the thermal contribution to scalar fluctuations. Furthermore, the scalar spectral index remains close to unity when the generalized slow-roll conditions are satisfied, indicating a nearly scale-invariant primordial scalar power spectrum.

We have also derived the tensor power spectrum and tensor-to-scalar ratio. Unlike scalar perturbations, tensor fluctuations are not directly sourced by the thermal bath. Consequently, thermal fluctuations enhance the scalar spectrum relative to the tensor spectrum, leading to a suppressed tensor-to-scalar ratio. The nonminimal coupling provides an additional mechanism for this suppression through its modifications of the effective background and perturbation coefficients. The resulting tensor-to-scalar consistency relation differs from that of standard cold inflation. Moreover, the modified temperature-to-Hubble ratio indicates that the warm-inflation condition $T>H$ can be maintained over suitable parameter ranges, including regimes with relatively weak dissipation. The modified field-excursion relation further demonstrates that sub-Planckian inflaton excursions can be accommodated in appropriate regions of parameter space.

In summary, our results demonstrate that nonminimal gravitational coupling, noncanonical kinetic dynamics, and thermal dissipation can act cooperatively to facilitate slow-roll evolution, enlarge the viable parameter space, and suppress primordial tensor perturbations. The proposed framework therefore provides a flexible generalization of conventional warm inflation, with the potential to accommodate nearly scale-invariant scalar perturbations and a small tensor-to-scalar ratio. Further investigations involving specific inflationary potentials, dissipation coefficients, and numerical constraints from cosmological observations will be valuable for assessing the phenomenological viability of this scenario in the future.

\acknowledgments

This work was supported by projects ZR2021MA037 supported by Shandong Provincial Natural Science Foundation, National Natural Science Foundation of China (Grant No. 11605100), the Natural Science Foundation of Henan under Grant No. 232300421351 and the Talent Introduction Fund (Grant No. 2020BS035) at Henan University of Technology.


%\begin{thebibliography}{99}

%\bibitem{a}
%Author, \emph{Title}, \emph{J. Abbrev.} {\bf vol} (year) pg.

%\bibitem{b}
%Author, \emph{Title},
%arxiv:1234.5678.

%\bibitem{c}
%Author, \emph{Title},
%Publisher (year).

% Please avoid comments such as "For a review'', "For some examples",
% "and references therein" or move them in the text. In general,
% please leave only references in the bibliography and move all
% accessory text in footnotes.

% Also, please have only one work for each \bibitem.
%\end{thebibliography}

\bibliographystyle{JCAP}
\bibliography{JCAP}
\end{document}